\documentclass[10pt,a4paper,twoside,english]{article}

\usepackage[T1]{fontenc}
\usepackage[latin9]{inputenc}
\usepackage{color}
\usepackage{float}
\usepackage{amsmath}
\usepackage{graphicx}
\usepackage{amssymb}
\usepackage[authoryear]{natbib}

\usepackage{babel}

\title{Weak Chaos and the ``Melting Transition'' in a Confined Microplasma
System}

\author{Chris Antonopoulos$^{1}$, Vasileios Basios$^{1}$ and Tassos Bountis$^{2}$}

\begin{document}

\maketitle

\begin{center}
$^{1}$Interdisciplinary Center for Nonlinear Phenomena and Complex
Systems (CeNoLi),
Service de Physique des Syst\`{e}mes Complexes et M\'{e}canique Statistique,\\
Universit\'{e} Libre de Bruxelles, 1050, Brussels,
Belgium\vspace{0.5cm}
\\
$^{2}$Center for Research and Applications of Nonlinear Systems
(CRANS),\\
Department of Mathematics,\\
University of Patras, 26500, Patras, Greece
\par\end{center}

\begin{abstract}
We present results demonstrating the occurrence of changes in the
collective dynamics of a Hamiltonian system which describes a
confined microplasma characterized by long--range Coulomb
interactions. In its lower energy regime, we first detect
macroscopically, the transition from a {}``crystalline--like'' to a
{}``liquid--like'' behavior, which we call the {}``melting
transition''. We then proceed to study this transition using a
microscopic chaos indicator called the \emph{Smaller Alignment
Index} (SALI), which utilizes two deviation vectors in the tangent
dynamics of the flow and is nearly constant for ordered
(quasi--periodic) orbits, while it decays exponentially to zero for
chaotic orbits as $\exp(-(\lambda_{1}-\lambda_{2})t)$, where
$\lambda_{1}>\lambda_{2}>0$ are the two largest Lyapunov exponents.
During the {}``melting phase'', SALI exhibits a peculiar,
stair--like decay to zero, reminiscent of {}``sticky'' orbits of
Hamiltonian systems near the boundaries of resonance islands. This
alerts us to the importance of the
$\Delta\lambda=\lambda_{1}-\lambda_{2}$ variations in that regime
and helps us identify the energy range over which {}``melting''
occurs as a multi--stage diffusion process through weakly chaotic
layers in the phase space of the microplasma. Additional evidence
supporting further the above findings is given by examining the
$GALI_{k}$ indices, which generalize SALI (=$GALI_{2}$) to the case
of $k>2$ deviation vectors and depend on the complete spectrum of
Lyapunov exponents of the tangent flow about the reference orbit.
\end{abstract}

\section{Introduction}

\label{intro}

It has long been established that microscopic deterministic chaos
provides an efficient mechanism for the mixing of orbits in the
phase--space of dynamical systems, leading to the decay of
statistical correlations as time evolves. Thus, chaotic dynamics can
magnify small scale fluctuations and justify the existence of
macroscopic variables like entropy and temperature, which are of
central importance in an analysis based on non--equilibrium
statistical mechanics \cite{gaspard-book}.

The main purpose of the work presented here is to study the
{}``melting transition'' in a microplasma model, using standard
methods, as well as certain recently--developed techniques for chaos
detection, such as the \emph{Smaller Alignment Index} (SALI)
\cite{kn:2,kn:4,kn:5} and its extension to the so--called
\emph{Generalized Alignment Index} (GALI) \cite{kn:3,kn:7,kn:10}.
This latter approach is based on geometrical aspects of the
microscopic dynamics and has significant computational advantages
over more classical indicators based either on local dynamics, such
as the Lyapunov exponents, or statistical properties, such as the
mean temperature or the Kolmogorov--Sinai entropy. In fact, the use
of these novel indices often constitutes an improvement of several
orders of magnitude in CPU and dynamical time for the identification
of the chaotic or ordered nature of single orbits
\cite{kn:14,kn:2,kn:4,kn:5,kn:7,kn:8,kn:9,kn:10}.

The utility of SALI (or GALI) methods in detecting dynamical
regime changes in few--particle Hamiltonian systems \cite{kn:7} is
due to their sensitivity in tracing out the geometrical properties
of the tangent dynamics of the flow. Thus, they provide accurate
information about regime changes when important parameters of the
system are varied. For example, the existence of {}``sticky''
regions and the occurrence of slow diffusion in weakly chaotic
domains is detectable and the distinction between order and
strongly chaotic motion is easily made by these methods
\cite{kn:7}. In the case of fully developed chaotic motion, SALI
is particularly efficient, since it decays exponentially as
$\exp(-(\lambda_{1}-\lambda_{2})t)$ where
$\lambda_{1}>\lambda_{2}>0$ are the two largest Lyapunov exponents
\cite{kn:5}, while for ordered orbits SALI$\;\propto\;$const.$>0$.

The system we consider here consists of $N$ interacting particles
described by a Hamiltonian function of the form \[
H(\vec{q},\vec{p})=K(\vec{p})+V(\vec{q})\]
 where the kinetic energy part $K(\vec{p})=\frac{1}{2}\sum_{i=1}^{N}p_{i}^{2}$
is quadratic in the generalized momenta
$\vec{p}=(p_{1},\ldots,p_{N})$ and the potential energy $V(\vec{q})$
is a function of its generalized position coordinates
$\vec{q}=(q_{1},\ldots,q_{N})$. In the case of confined systems, one
takes $V$ as the sum of two terms:
$V(\vec{q})=V_{tr}(\vec{q})+V_{in}(\vec{q})$, where
$V_{tr}(\vec{q})$ represents the potential of the trap (being a
smooth positive function), while $V_{in}(\vec{q})$ accounts for the
interactions amongst the $N$ particles. The fact that $N$ is a
finite number classifies the system as {}``small'', in contrast to a
{}``large'' thermodynamic system (where one lets $N$ and the volume
$V$ tend to infinity in such a way that $N/V$ is a finite real
constant) \cite{Hill-1994,Hill56}.

Small, finite Hamiltonian systems in contrast to large, infinite,
systems do not exhibit phase transitions in the standard sense.
The characteristic discontinuities or singularities in
thermodynamic functions or their derivatives, which is the
signature of a phase transition in large systems would now appear
as steep but continuous changes in their thermodynamic functions
or their derivatives \cite{Hill-1994}. 

Phase transitions in small systems, have also been associated, quite recently, with certain
topology changes in configuration space \cite{casetti, Casetti2000237}.
These developments are stemming from earlier work on Hamiltonian systems 
exhibiting chaotic instabilities associated with singular behaviour
of their configuration--space curvature fluctuations at 
their phase transition point \cite{PhysRevA.36.9620, PhysRevLett.79.4361,PhysRevE.56.2508}.
Furthermore, investigations of the temperature dependence of the largest Lyapunov 
exponent and other observables related to the ``topological hypothesis'' and the issue of phase transitions 
in many--degrees of freedom systems is thoroughly presented in \cite{Casetti2000237} and references therein.

In this paper, our model Hamiltonian system describes a microplasma
characterized by long range (non shielded) Coulomb interactions
described by the potential $V_{in}(\vec{q})$. The microplasma is
confined in a Penning trap given by the potential $V_{tr}(\vec{q})$
and our aim is to study the motion of its ions as the energy
increases. The system evolves, in general, in 3 spatial dimensions
and contains a relatively small number of ions for which the
tracking of individual trajectories is numerically as well as
experimentally feasible \cite{kn:26,kn:27,kn:28,kn:29,kn:30,kn:31}.

It is important to note that Lyapunov exponents have already been
used to investigate spatially extended plasmas
\cite{fromPierre26,fromPierre27,fromPierre28} as well as
1--dimensional wave--particle plasma models
\cite{fromPierre29,fromPierre30}. Moreover, as Gaspard \cite{kn:1}
has demonstrated, for a realistic Hamiltonian model in 3 spatial
dimensions containing a relatively small number of ions, the long
range nature of the Coulomb interaction $V_{in}(\vec{q})$ makes
the maximum Lyapunov exponents behave differently in microplasmas
than in many--particle systems with short--range interactions,
such as a hard--ball fluid or systems with shielded ionic
interactions (Yukawa potentials). A detailed analysis of the
differences between shielded (Yukawa--like) and unshielded Coulomb
potentials can be found in the book by Baus and Tejero
\cite{kn:100}. Indeed, in microplasmas composed of more than a few
dozen of ions, the behavior of the maximum Lyapunov exponent,
especially in the ``gas phase'', was explained following purely
statistical mechanics arguments \cite{kn:1}.

The plan of the paper is the following: The Hamiltonian
representation of our system and a summary of previous work are
presented in Sec. 2. Some background material related to the SALI
method and the spectrum of Lyapunov exponents are given in Sec. 3.
The calculation of Lyapunov exponents spectra for our system is
presented in Sec. 4, while the detection of the ``melting
transition'' as a passage from weak to strong chaos is
demonstrated in Sec. 5. Finally, our conclusions are presented in
Sec. 6.

\section{Description of the model and summary of previous work\label{model_description}}

Let us consider a microplasma of $N$ ions of equal mass $m=1$ and
electric charge $q$ in a Penning trap with electrostatic potential
\begin{equation}
\Phi(x,y,z)=V_{0}\frac{2z^{2}-x^{2}-y^{2}}{r_{0}^{2}+2z_{0}^{2}}\end{equation}
 and constant magnetic field along the $z$ direction with a vector
potential of the form \begin{equation}
\mathbf{A}(x,y,z)=\frac{1}{2}(-By,Bx,0).\end{equation}

Then, the Hamiltonian of the full system is given by
\begin{equation}
\mathcal{H}=\sum_{i=1}^{N}\Biggl\{\frac{1}{2m}(\mathbf{p}_{i}-q\mathbf{A}(\mathbf{r}_{i}))^{2})+q\Phi(\mathbf{r}_{i})\Biggr\}+\sum_{1\leq
i<j\leq
N}\frac{q^{2}}{4\pi\epsilon_{0}r_{ij}}\label{initial_Hamiltonian}\end{equation}
 where $\mathbf{r_{i}}$ is the position of the $i$th ion, $r_{ij}$
is the Euclidean distance between the $i$th and $j$th ions and $\epsilon_{0}$
is the vacuum permittivity. In the Penning trap, the ions are subjected
to a harmonic confinement in the $z$ direction with frequency \begin{equation}
\omega_{z}=\sqrt{\frac{4qV_{0}}{m(r_{0}^{2}+2z_{0}^{2})}}\end{equation}
 while in the perpendicular direction (due to the cyclotron motion)
they rotate with frequency $\omega_{c}=qB/m$. Thus, in a frame
rotating around the $z$ axis at the Larmor frequency
$\omega_{L}=\omega_{c}/2$, the ions feel a harmonic confinement of
frequency
$\omega_{x}=\omega_{y}=\sqrt{\frac{\omega_{c}^{2}}{4}-\frac{\omega_{z}^{2}}{2}}$
in the direction perpendicular to the magnetic field. In the
rescaled time $\tau=\omega_{c}t$, position
$\mathbf{R}=\mathbf{r}/a$ and energy
$H=\frac{\mathcal{H}}{m\omega_{c}^{2}a^{2}}$ with
$a=\Bigl({\frac{q^{2}}{4\pi\epsilon_{0}m\omega_{c}^{2}}}\Bigr)^{\frac{1}{3}}$,
the Hamiltonian \eqref{initial_Hamiltonian} describing the motion
takes the form \begin{equation}
H=\sum_{i=1}^{N}\Bigl[\frac{1}{2}\mathbf{P}_{i}^{2}\Bigr]+\sum_{i=1}^{N}\Bigl[\Bigl(\frac{1}{8}-\frac{\gamma^{2}}{4}\Bigr)(X_{i}^{2}+Y_{i}^{2})+\frac{\gamma^{2}}{2}Z_{i}^{2}\Bigr]+\sum_{i<j}\frac{1}{R_{ij}}=E\label{mic_plas_Ham}\end{equation}
 where $E$ is the total energy of the system, $\mathbf{R}_{i}=(X_{i},Y_{i},Z_{i})$
and $\mathbf{P}_{i}=(P_{X_{i}},P_{Y_{i}},P_{Z_{i}})$ are the positions
and canonically conjugate momenta respectively, $R_{ij}$ is the Euclidean
distance between different ions $i,j$ given by \begin{equation}
R_{ij}=\sqrt{(X_{i}-X_{j})^{2}+(Y_{i}-Y_{j})^{2}+(Z_{i}-Z_{j})^{2}}\end{equation}
 and $\gamma=\omega_{z}/\omega_{c}$.

The ions are trapped in bounded motion under the condition that \begin{equation}
0<|\gamma|<\frac{1}{\sqrt{2}}.\end{equation}
 The trap is called prolate if $0<|\gamma|<\frac{1}{\sqrt{6}}$, isotropic
if $|\gamma|=\frac{1}{\sqrt{6}}$ and oblate if
$\frac{1}{\sqrt{6}}<|\gamma|<\frac{1}{\sqrt{2}}$. So, the motion
is quasi 1--dimensional in the limit $\gamma\rightarrow0$ and
quasi 2--dimensional in the limit $\gamma\rightarrow1/\sqrt{2}$.
The $Z$ direction is a symmetry axis and hence the $Z$ component
of the angular momentum
$L_{Z}=\sum_{i=1}^{N}X_{i}P_{Y_{i}}-Y_{i}P_{X_{i}}$ is conserved,
being thus, a second integral of the motion. We suppose from now
on that the angular momentum is equal to zero (i.e. $L_{Z}=0$) and
that the motion is studied in the Larmor rotating frame.

In \cite{kn:1}, it was shown that the motion of the ions governed by
Hamiltonian \eqref{mic_plas_Ham} is generally very sensitive to
initial conditions and has at least one positive maximum Lyapunov
exponent, while a first study of its dependence on the energy,
number of ions and trap geometry was also presented. At low kinetic
energies, where the microplasma forms an ion crystal, chaos is
considerably reduced, since the motion is quasi--harmonic around
stable equilibrium configurations. On the other hand, at high
temperatures (or kinetic energies), the maximum Lyapunov exponent
decreases, as Coulomb interactions become negligible and the
microplasma forms a thermal cloud of nearly independent ions moving
throughout the full extent of the harmonic potential of the trap.
Finally, for intermediate values of the energy, there is a regime of
chaotic behavior which becomes wider as the number of ions
increases. It is in this regime that the maximum Lyapunov exponent
attains a peak, whose value increases as a function of the number of
ions.

A crucial aspect related to our work is that of the geometry of
the trap. As has been demonstrated in \cite{kn:1} for prolate
traps the maximum Lyapunov exponent (and therefore the K-S entropy
of the system) show a distinctively smoother increase in
comparison to the oblate traps in the energy values interval from
$E_{min}$ (its minimum) to $E_{0}$ (its maximum). The transition
of interest for our analysis is taking place within a small
interval of energies at the beginning, i.e. just after $E_{min}$.
Therefore, the slower the increase of the function $E=E(H)$ the
higher the resolution required and the more accurate (and
time--consuming) the numerical computation needed. Of course, if
one argues in analogy to the standard phase transition theory, the
dimensionality of the problem makes it easier to detect
transitions in two or three spatial dimensions. 
But systems with long range interactions may still exhibit a
transition even in one dimension. This justifies the choice of
small $\gamma$ adopted in the present paper as being both
physically challenging and numerically tractable.

\subsection{Crossover through different dynamical regimes}

At zero temperature, the system freezes at a crystalline state (see
\cite{kn:1} and in particular refs. {[}20{]} -- {[}25{]} therein)
reminiscent of Wigner crystals \cite{0034-4885-70-12-R02}. In that
case, the ion crystal is composed of several concentric rings for
the case of an oblate (quasi 2--dimensional) trap or a single line
of ions in the case of a prolate (quasi 1--dimensional) trap. As
temperature rises slightly above zero, the ions execute
quasi--periodic motion about stable periodic orbits moving around
their equilibrium position. This is the regime where stable normal
modes of vibration play an important role, as expected.

At slightly higher temperatures, the system ``bifurcates'' from
the regime of quasi--harmonic motion to one where it executes
quasiperiodic motion in the form of a collective soft-mode. We use
the term ``soft-mode'' here since the dynamics is similar to what
would be expected under a collective force on the particles, which
diminishes as the energy increases. The frequencies of the normal
vibrational modes become smaller and smaller and tend to zero at
the transition.

As the energy continues to increase, these soft modes overcome the
energy barriers and set out to explore broader domains of phase space.
At even higher temperatures, their motion becomes erratic and the
ion crystal ``melts'', entering a regime of strongly chaotic motion
in phase space.

As the temperature increases further, the ions form a thermal
cloud in which the mean Coulomb potential energy starts to become
negligible with respect to the mean kinetic energy and mean
harmonic potential energy. It has been shown that the maximum
Lyapunov exponent $\lambda_{1}$ reaches its peak value at energies
well above the ``melting phase'', while this transition is not
manifested in the behavior of the Lyapunov exponents as a function
of the number of particles $N$ or the energy $E$.

Finally, for energies beyond the peak value of the maximum
Lyapunov exponent, the motion is dominated by the kinetic energy
part and the system becomes amenable to a statistical mechanical
treatment with considerable accuracy, in spite of the small number
of particles present. Indeed, at such high temperatures $T$, the
spatial disorder of the microplasma can be described in terms of
its thermodynamic entropy and its maximum Lyapunov exponent
$\lambda_{1}$ expressed by the theoretical estimate \cite{kn:1}
\begin{equation} \lambda_{1}\sim\left\langle
\frac{N^{2}}{R_{ij}}\right\rangle ^{\frac{1}{2}}\sim N\frac{(\ln
T)^{\frac{1}{2}}}{T^{\frac{3}{4}}}\;\mbox{for}\;
T\rightarrow\infty\end{equation}
 which may be used to characterize the crossover to the regime of
thermal cloud motion, albeit at a high computational cost for the
determination of $\lambda_{1}$.

\begin{figure}[H]

\begin{centering}
\includegraphics[width=1\textwidth,height=0.5\textwidth]{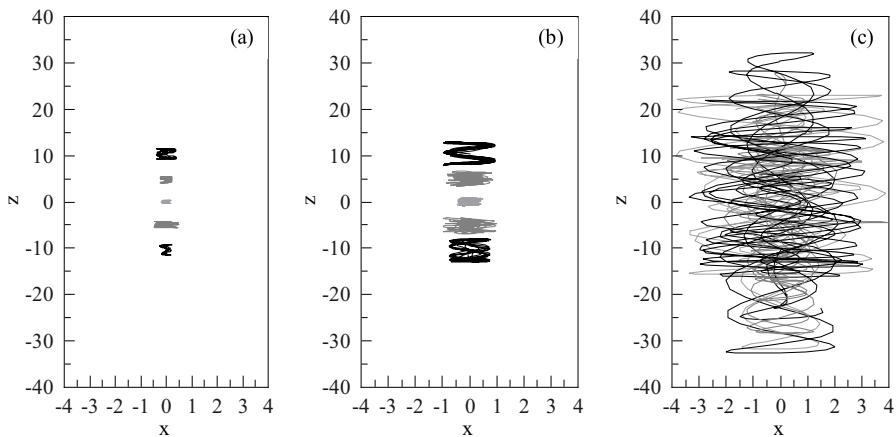}
\par\end{centering}

\caption{Plot of typical orbits for $N=5$ and $\gamma=0.07$ (prolate quasi
1--dimensional geometry): (a) At $E=2$ the motion is dominated by
its stable normal modes and {}``crystalline'' behavior is observed.
(b) At $E=2.35$ a transition to chaotic motion occurs, where {}``melting''
is expected to take place. (c) At $E=9$ the onset of {}``thermal
cloud'' behavior is evident.\label{fig:4} }

\end{figure}

Furthermore, in \cite{kn:1}, the maximum value of $\lambda_{1}$ is
numerically observed to obey a power law with respect to the
energy as well as the number of ions. Thus, one reaches the
conclusion that, for a prolate trap geometry (quasi
1--dimensional), the {}``melting transition'' and its associated
microscopic dynamical behavior should be detectable at low
energies and few number of ions (e.g. for $N=5$ and
$\gamma=0.07$).

The different characteristic kinds of motion in such a setting are
depicted in Fig. \ref{fig:4}. Panel (a) presents a typical
behavior of the ions at energy $E=2$, where the dynamics is
dominated by stable normal modes that correspond to
{}``crystalline'' behavior. At the slightly increased energy
$E=2.35$ of panel (b) we observe the occurrence of a transition
from stable collective (quasi--periodic) motion to chaotic
behavior. Finally, in panel (c), the onset of a fully
{}``thermal'' cloud behavior is apparent at the energy $E=9$. So,
for the rest of the paper we consider the case of $N=5$ ions in
the prolate trap geometry with $\gamma=0.07$.

\begin{figure}[H]

\begin{centering}
\includegraphics[width=1\textwidth,height=0.5\textwidth]{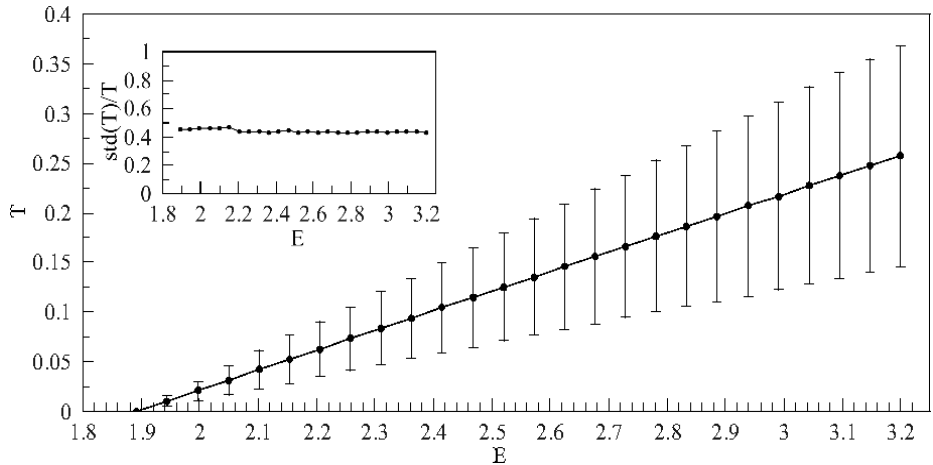}
\par\end{centering}

\caption{Plot of the dependence of the temperature $T$ (as in eq. \eqref{Tdef}, dimensionless scale)
on the energy $E$ for the microplasma system \eqref{mic_plas_Ham}
of $N=5$ ions in a prolate Penning trap $(\gamma=0.07)$. The error
bars represent one standard deviation from the time averaged temperature $T$. Clearly,
the crossover from ordered to weakly chaotic motion around $E\in(2,2.5)$
does not occur in a manner justifying the use of statistical mechanics
considerations. In the inset, the ratio of the standard deviation of the time averaged $T$ 
to the time averaged temperature $T$ shows that they are comparable.\label{temper}}

\end{figure}

The temperature, in dimensionless units, 
\begin{equation}\label{Tdef}
T\propto v_{rms}^2=\left\langle\Vert\mathbf{P}\Vert^2\right\rangle_N\equiv\frac{1}{N}\sum_{i=1}^{N}\Vert\mathbf{P}_i\Vert^2 
\end{equation}
(where $\Vert\mathbf{P}_i\Vert$ denotes the Euclidean norm of the $i$th ions' velocity $\mathbf{P}_i$) averaged over the time $t$ 
which is a proportional quantity of the mean kinetic energy of the ions is plotted in Fig. \ref{temper}, as a function of the energy $E$ for the microplasma \eqref{mic_plas_Ham} with $N=5$ ions in the prolate Penning trap with $\gamma=0.07$. Evidently, the melting of the crystal here is
not associated with a sharp increase of $T$ at some critical
energy, as would be expected from a first order phase transition.
Note also in Fig. \ref{temper} that the fluctuations of $T$ (measured
by its standard deviation) cannot provide a clear indication of the
crossover from a regime of stable oscillations, through {}``melting'',
to strongly developed chaotic motion. So, the use of statistical mechanics
considerations must focus on the importance of large fluctuations.
Indeed, the inset of Fig. \ref{temper} shows that the fluctuations
of the time averaged temperature $T$, in the crossover regime, are comparable with the values of the time averaged temperature itself.

One can, therefore, say that there is no {}``macroscopic'' methodology
for detecting dynamical regime changes in system \eqref{mic_plas_Ham}
that might be useful to theorists or experimentalists. It, therefore,
becomes especially important to adopt a different approach and attempt
to study dynamical regime changes in our microplasma system, by performing
a detailed study of its \textit{microscopic} dynamics.

To this end, we shall focus hereafter on the lower energy range of
Hamiltonian \eqref{mic_plas_Ham}, where the transition from
{}``crystalline--like'' to {}``liquid--like'' collective behavior
takes place. In the next section, we review the methods used for
the detection of these dynamical regimes, i.e. the Smaller
Alignment Index (SALI) \cite{kn:2,kn:4,kn:5} and the spectrum of
Lyapunov exponents \cite{kn:11,kn:12} and compare the information
provided by these approaches.

As we shall see, in the crossover regime, the SALI exhibits a rather
intricate {}``stair--like'' power law decay to zero, with different
Lyapunov exponents, reminiscent of the {}``stickiness'' behavior
typically observed in the neighborhood of resonance islands of
Hamiltonian systems \cite{kn:7}. In our case, this type of weakly
chaotic dynamics will turn out to characterize the energy range over
which the passage from ordered to irregular dynamics occurs, through
what we call the {}``melting transition''.

\section{Methods for distinguishing between order and chaos}

\label{background}

\subsection{The Smaller Alignment Index (SALI)}

\label{SALI}

The SALI method was initially introduced in \cite{kn:2} and has been
applied successfully to distinguish between ordered and chaotic
dynamics in maps of various dimensions \cite{kn:15,kn:14}, in
Hamiltonian systems \cite{kn:4,kn:5}, as well as in problems of
celestial mechanics \cite{kn:20,kn:21}, galactic dynamics
\cite{kn:17}, field theory \cite{kn:16} and nonlinear 1--dimensional
lattices \cite{kn:9,kn:8,kn:19}.

Following \cite{kn:2}, one considers the $2N$--dimensional phase
space of an arbitrary autonomous Hamiltonian system \begin{equation}
H(\vec{q},\vec{p})\equiv
H(q_{1}(t),\ldots,q_{N}(t),p_{1}(t),\ldots,p_{N}(t))=E\label{1}\end{equation}
 where $\vec{q}$ is the canonical position vector, $\vec{p}$ is
the corresponding canonical conjugate momentum vector and $E$ is
the total constant energy. The time evolution of an orbit of Hamiltonian
\eqref{1} associated with an initial condition $\vec{x}(t_{0})=(q_{1}(t_{0}),\ldots,q_{N}(t_{0}),p_{1}(t_{0}),\ldots,p_{N}(t_{0}))$
at time $t=t_{0}$ is defined as the solution of Hamilton's equations
of motion \begin{equation}
\frac{dq_{i}(t)}{dt}=\frac{\partial H}{\partial p_{i}(t)},\;\frac{dp_{i}(t)}{dt}=-\frac{\partial H}{\partial q_{i}(t)},\; i=1,\ldots,N\label{2}\end{equation}
 and is called the orbit $\vec{x}(t)$ of eqs. \eqref{2} passing
through $\vec{x}(t_{0})$.

To define the Smaller Alignment Index (SALI), one uses the \emph{variational
equations}, which represent the linearization of Hamilton's equations
of motion \eqref{2} about a reference orbit $\vec{x}(t)$ of the
system and are defined as \begin{equation}
\frac{d\vec{\upsilon_{i}}(t)}{dt}=\mathcal{J}(\vec{x}(t))\cdot\vec{\upsilon_{i}}(t),\,\,\,\,\forall i=1,\ldots,2N\label{3}\end{equation}
 where $\mathcal{J}(\vec{x}(t))$ is the Jacobian of the right--hand
side of eqs. \eqref{2} calculated about the orbit $\vec{x}(t)$.
The vectors $\vec{\upsilon_{i}}(t),\forall i=1,\ldots,2N$ are known
as deviation vectors of the $\vec{x}(t)$. If we follow two of them,
say $\vec{\upsilon_{k}}(t)$ and $\vec{\upsilon_{l}}(t)$, we can
define the \emph{Smaller Alignment Index} (SALI) as

\begin{equation}
\mbox{SALI}(t)\equiv\min\bigg\{\bigg\|\frac{\vec{\upsilon_{k}}(t)}{\|\vec{\upsilon_{k}}(t)\|}-\frac{\vec{\upsilon_{l}}(t)}{\|\vec{\upsilon_{l}}(t)\|}\bigg\|,\;\bigg\|\frac{\vec{\upsilon_{k}}(t)}{\|\vec{\upsilon_{k}}(t)\|}+\frac{\vec{\upsilon_{l}}(t)}{\|\vec{\upsilon_{l}}(t)\|}\bigg\|\bigg\}\label{5}\end{equation}
 where $\|\cdot\|$ denotes the usual Euclidean norm defined in $\mathbb{R}^{2N}$.

If $\vec{x}(t)$ is chaotic then
$\lim_{t\rightarrow\infty}\mbox{SALI}(t)=\min\{0,2\}=0$ since both
deviation vectors tend to align with the direction of the maximum
Lyapunov exponent as $t$ increases \cite{kn:5,kn:13}. Furthermore,
it has been shown that the time evolution of the SALI for chaotic
orbits tends to zero exponentially at a rate related to the
difference of the two largest Lyapunov exponents $\lambda_{1}$ and
$\lambda_{2}$ as \cite{kn:5} \begin{equation}
\mathrm{SALI}(t)\propto
e^{-(\lambda_{1}-\lambda_{2})t}.\label{eq:SALI-exp}\end{equation}
 On the other hand, if the orbit is ordered, SALI exhibits small oscillations
about a constant $\alpha\in(0,\sqrt{2}]$. This is so, because both
deviation vectors tend to become tangential to the torus on which
the orbit is evolving while remaining mutually linearly independent
in time due the existence of local or global integrals of motion
\cite{kn:4}.

It follows that this sharply different behavior of the SALI for
ordered and chaotic orbits makes it a reliable and computationally
fast tool for distinguishing between order and chaos in
Hamiltonian systems of any number of degrees of freedom. The
choice of the initial deviation vectors is arbitrary and in
general does not affect the method apart from some very special
cases demonstrated theoretically and verified numerically in a
number of previous works \cite{kn:2,kn:4,kn:5}.

\subsection{The spectrum of Lyapunov exponents}

\label{Lyap_exps}

One of the most standard and well--established means for extracting
information about the nature of a given orbit of a dynamical system
is to calculate its maximum \emph{Lyapunov Characteristic Exponent}
(LCE) $\lambda_{1}$. If $\lambda_{1}>0$ the orbit is characterized
as chaotic. The theory of Lyapunov exponents was first applied to
characterize chaotic orbits by Oseledec \cite{kn:22}, while the
connection between Lyapunov exponents and exponential divergence of
nearby orbits was given in \cite{kn:23,kn:24}. Benettin et al.
\cite{kn:11,kn:12} studied the problem of the computation of all
LCEs theoretically and proposed in \cite{kn:12} an algorithm for
their efficient numerical evaluation. In particular, $\lambda_{1}$
is computed as the limit for $t\rightarrow\infty$ of the quantity
\begin{equation}
L_{1}(t)=\frac{1}{t}\,\ln\frac{\|\vec{\upsilon_{1}}(t)\|}{\|\vec{\upsilon_{1}}(0)\|}\,,\,\mbox{i.e.}\;\lambda_{1}=\lim_{t\rightarrow\infty}L_{1}(t)\label{eq:lyap1_def}\end{equation}
 where $\vec{\upsilon_{1}}(0)$, $\vec{\upsilon_{1}}(t)$ are deviation
vectors from the orbit we want to characterize, at times $t=0$ and
$t>0$ respectively. Similarly, all other LCEs,
$\lambda_{2}$,$\lambda_{3},\ldots,\lambda_{2N}$ are computed as
limits for $t\rightarrow\infty$ of analogous quantities,
$L_{2}(t)$,$L_{3}(t),\ldots,L_{2N}(t)$. In the present paper, we
shall compute the values of all Lyapunov exponents (called the
Lyapunov spectrum) using the algorithm proposed by Benettin et al.
\cite{kn:11,kn:12}.

Let us also recall that the Lyapunov spectrum of an $N$ degree of
freedom Hamiltonian system, i.e. $\{\lambda_{i}\}_{i=1}^{2N}$, exhibits
a basic symmetry with $N-1$ of its members having the opposite sign
of the other $N-1$ and two exponents, i.e. $\lambda_{N}$, $\lambda_{N+1}$,
being equal to 0. So, the discussion from here on concerns only the
positive half of the Lyapunov spectrum, $\{\lambda_{i}\}_{i=1}^{N-1}$
where $\lambda_{i}>0$.

The calculation of the Lyapunov spectrum is computationally demanding,
as its convergence often requires the integration of trajectories
over very long time intervals. This convergence depends on the inverse
of the largest Lyapunov exponent or \emph{Lyapunov time}. So, the
closer $\lambda_{1}$ is to zero the longer the integration is needed
to obtain reliable estimates for the full spectrum.

On  the other hand, of particular importance to statistical
mechanics is the connection of the sum of all positive Lyapunov
exponents (an index of chaos based on microscopic quantities) to
the Kolmogorov--Sinai entropy $H_{KS}$ (a measure of disorder of a
macroscopic nature), given by Pesin's celebrated theorem
\cite{kn:24} \begin{equation}
H_{KS}=\sum_{\lambda_{i}>0}\lambda_{i}\label{eq:Pesin}\end{equation}
 in the case of zero escape rates. This connection has opened new
directions in the applications of thermodynamics, using the
underlying microscopic dynamics of chaotic orbits (for a review see
\cite{DasBook,gaspard-book}).

\section{Calculation of Lyapunov spectra}

\label{LES_calc}

\begin{figure}[H]

\begin{centering}
\includegraphics[width=1\textwidth,height=0.5\textwidth]{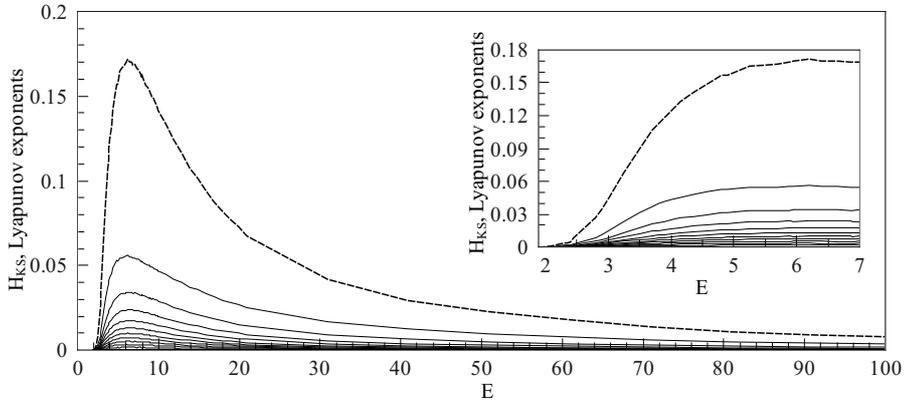}
\par\end{centering}

\caption{The Kolmogorov--Sinai entropy $H_{KS}$, for $\gamma=0.07$ and $N=5$, 
computed by the eq. \eqref{eq:Pesin}
(dashed highest line) and the Lyapunov exponents (black solid lines)
are plotted from the maximum down to the lowest one (15th) as a function
of the energy. Note that, as shown in the inset, their maximum occurs
at energies close to $E\approx6.2$, which is much higher than the
regime where the {}``melting transition'' takes place (see text).}

\label{fig:1}
\end{figure}

For the purposes of simplifying our analysis, let us consider the
case of few ions, e.g. $N=5$ and $\gamma=0.07$ (i.e. a small system
in a prolate trap). As mentioned above, in this setting, it is
easier to detect dynamical regime changes, such as seen in Fig.
\ref{fig:1}, where we have plotted the Kolmogorov--Sinai entropy
$H_{KS}$ (see eq. \eqref{eq:Pesin}) as a function of the energy $E$
of the microplasma Hamiltonian \eqref{mic_plas_Ham}. Both $H_{KS}$
and Lyapunov exponents grow less steeply than in the oblate case
(quasi 2--dimensional trap) for the region of small energies, as
already pointed out in \cite{kn:1}.

This slow increase of $H_{KS}$ and $\lambda_{i}$'s exhibits an
inflection point at low energies as is evident in the inset of
Fig. \ref{fig:1}. We remark that this inflection point occurs just
above the energy threshold that permits ions to move around their
fixed points, while beyond this energy we observe the transition
from weak to strong chaotic behavior. Note that $H_{KS}$ as well
as the Lyapunov exponents
exhibit a maximum around $E\approx6.2$. %

\begin{figure}[H]
 \begin{raggedright} \includegraphics[width=0.5\textwidth,height=0.5\textwidth]{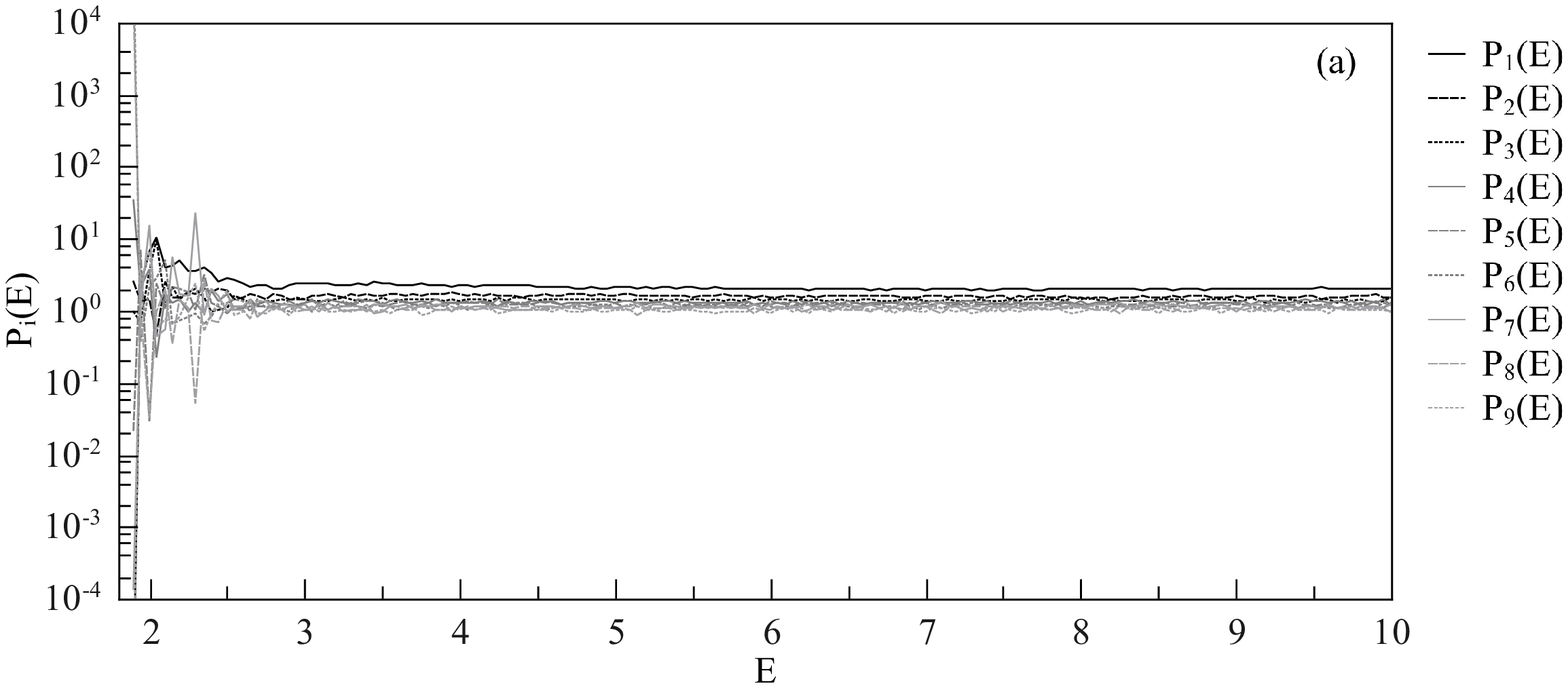}\includegraphics[width=0.5\textwidth,height=0.5\textwidth]{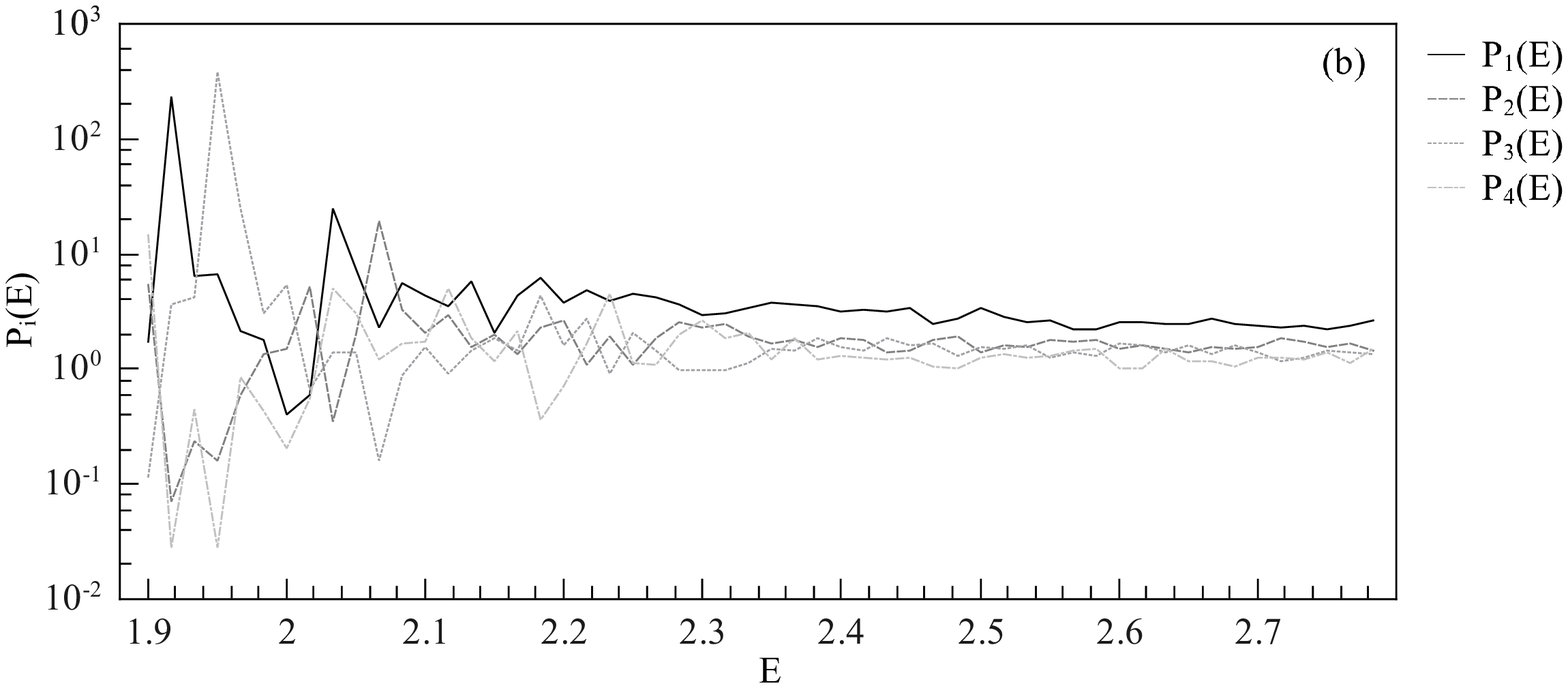}

\end{raggedright}

\caption{(a) Plot of the logarithms of the ratios ${\cal P}_{i}(E),\; i=1,\ldots,9$
(see eq. \eqref{LES_ratios}), for $\gamma=0.07$ and $N=5$, as a function of the energy $E$. Note
the characteristic gaps in the spectrum appearing for large values
of $E$. (b): Same as in panel (a) for $i=1,\ldots,4$ and small values
of $E$, with the characteristic gaps in the spectrum becoming erratic
in the regime of weak chaos, settling down to more clearly defined
values around $E\geq2.3$.}

\label{fig:2}
\end{figure}

As a first attempt to derive useful information from the spectrum
of Lyapunov exponents (see Fig. \ref{fig:1}), in the energy range
where the {}``melting transition'' is expected to occur, we
calculate the logarithms of the ratios of successive pairwise
differences of Lyapunov exponents
\begin{equation} {\cal P}_{i}(E)=\frac{\lambda_{i}-\lambda_{i+1}}{\lambda_{i+1}-\lambda_{i+2}},\;
i=1,\ldots,N-2,\;\lambda_{i}\equiv\lambda_{i}(E)\label{LES_ratios}\end{equation}
 as a function of the energy (see Fig. \ref{fig:2}). The calculation
of all $\lambda_{i}$'s was carried out for each trajectory up to
a final integration time $t_{f}=10^{6}$, ensuring that a relative
convergence to 4 or 5 significant decimal digits has been achieved.

Note in Fig. \ref{fig:2}(a) that the ratios (\ref{LES_ratios})
fluctuate wildly initially and only for $E\geq2.3$ do they start
to settle down to small oscillations about their ultimate values,
which, at least for the first 2 or 3 ratios appear to be quite
distinct. This suggests that it is within an interval
$E_{0}\lesssim E<2.3$ where $E_{0}\simeq1.89$ that we should
expect to find the {}``melting transition'', where the positive
Lyapunov exponents are very small, far from the values they attain
in the regime of strong chaos (see inset of Fig. \ref{fig:1}
for $E\geq2.3$). %

Undoubtedly, although the information concerning the existence of
two distinct dynamical regimes at small energies is somehow
contained in the plots of Fig. \ref{fig:2}, expressions
\eqref{LES_ratios} still fail to provide a precise picture of this
change. This is perhaps due to inaccuracies of the inherent
averaging process used in the calculation of the ${\cal P}_{i}(E)$ at low energies, where the Lyapunov exponents are very
small. We, therefore, proceed to study more carefully the
underlying local microscopic dynamics by employing the SALI method
we introduced previously.

\section{Weak chaos detection}

\label{weak_chaos}

In this section, we apply the procedure outlined in Sec. 3.1 to
analyze microscopically trajectories of the few--particle (i.e.
$N=5$ ions) microplasma system described by the Hamiltonian
\eqref{mic_plas_Ham} in the prolate trap with $\gamma=0.07$. We
are thus able to identify at low energies a regime of weak chaos,
through which the system passes from a {}``crystalline--like''
type of ordered motion to the strongly chaotic behavior of
collective soft modes. This is achieved by detecting qualitative
changes in the SALI behavior for trajectories within the energy
interval where the transition occurs.

\begin{figure}[H]

\begin{centering}
\includegraphics[width=0.5\textwidth,height=0.4\textwidth]{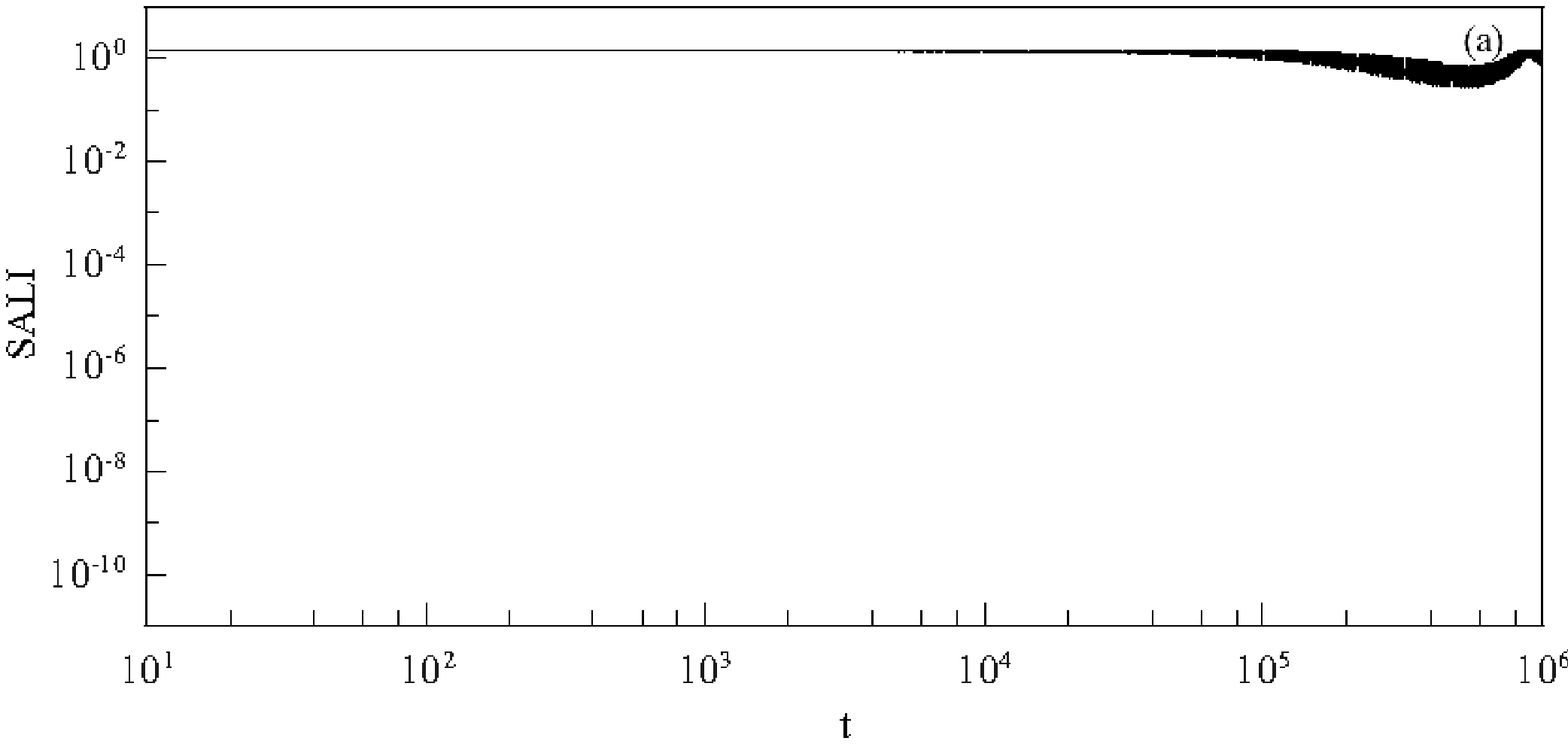}\includegraphics[width=0.5\textwidth,height=0.4\textwidth]{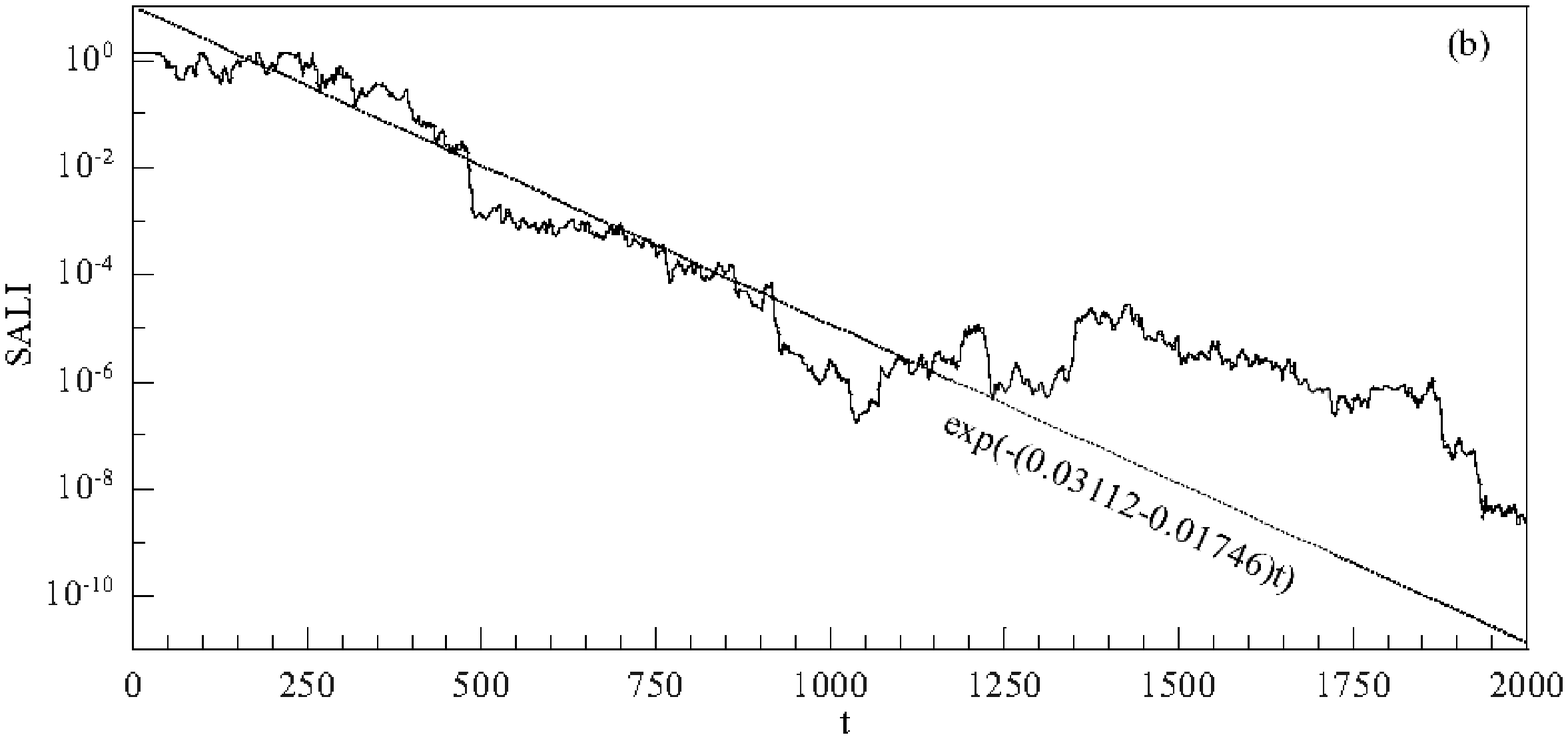}
\par\end{centering}

\begin{centering}
\includegraphics[width=0.5\textwidth,height=0.4\textwidth]{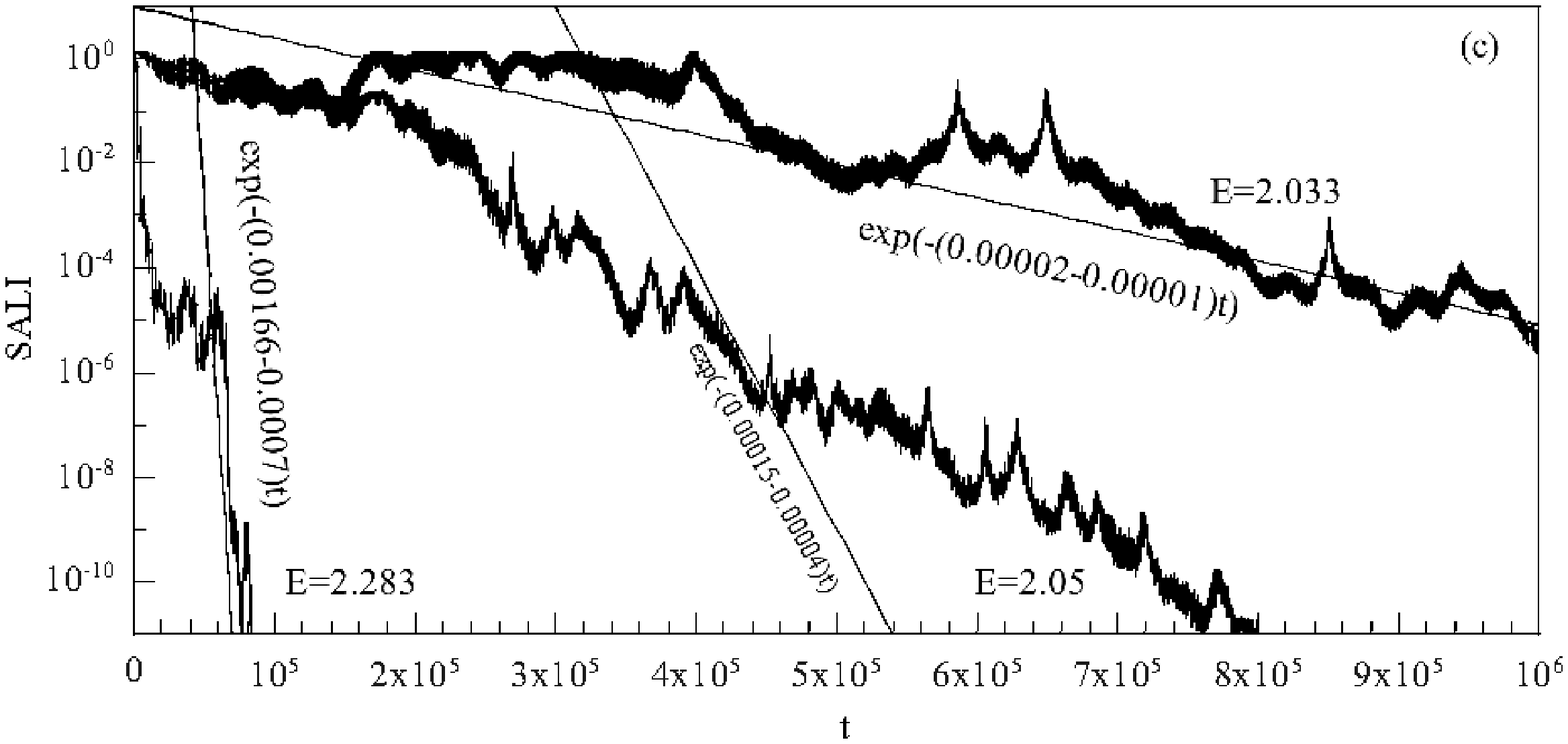}
\par\end{centering}

\caption{Plot of the evolution of SALI as a function of time $t$ for $N=5$,
$\gamma=0.07$ and 5 typical energies: (a) $E=1.9$ where the motion
is ordered. (b) $E=3.5$ (strong chaotic behavior) where we also plot
the theoretical prediction SALI$(t)\propto e^{-(\lambda_{1}-\lambda_{2})t}$
for comparison. (c) $E=2.033,\; E=2.05$ and $E=2.283$, plotting
also the corresponding theoretical predictions. Note that all three
vertical axes are logarithmic. \label{fig:3}}

\end{figure}

Fig. \ref{fig:3} shows the behavior of the SALI as a function of
time $t$ at selected number of energies. These plots represent
three distinct classes of long time behaviors for typical
trajectories within the interval $1.9\lesssim E\leq3.5$. We have
also computed, for the same sample of energies, the corresponding
Lyapunov spectra for sufficiently long integration times
(typically $t_{f}=10^{6}$) to make sure that the Lyapunov
exponents have converged to their limiting values up to a desired
accuracy (typically to 4 or 5 significant decimal digits).

Fig. \ref{fig:3}(a) represents the SALI evolution for a
representative trajectory of system \eqref{mic_plas_Ham} at
$E=1.9$. We see that it fluctuates around non--zero positive
values up to an integration time of order $t_{f}=10^{6}$,
indicating that the motion (at least up to that time) is ordered
and quasi-periodic. By contrast, in Fig. \ref{fig:3}(b), we follow
an orbit with energy $E=3.5$ and find that SALI decays to zero
exponentially fast, driven by the first gap in the Lyapunov
spectrum of Fig. \ref{fig:2}, according to the formula
$e^{-(\lambda_{1}-\lambda_{2})t}$ \cite{kn:5}. In this case, the
first two Lyapunov exponents of the trajectory,
$\lambda_{1}\approx0.03112$ and $\lambda_{2}\approx0.01746$, are
large enough and distinct, implying strongly chaotic motion.

Finally, let us turn our attention to the intermediate and more
interesting case where SALI exhibits a stair--like decay to zero
as a function of time, at energies $2<E<2.3$ (see Fig.
\ref{fig:3}(c)). Choosing, for example, a trajectory with
$E=2.033$ we observe that SALI differs significantly compared to
what is shown in Fig. \ref{fig:3}(a) or \ref{fig:3}(b). At first,
one might think that, as in Fig. \ref{fig:3}(a), the orbit is
quasi-periodic, since SALI {}``sticks'' to non--zero values for a
fairly long time interval ($t\lesssim4.4\times10^{5}$). Then, its
behavior changes qualitatively, showing a near exponential decay
to zero $\propto e^{-(\lambda_{1}-\lambda_{2})t}$. This suggests
that, even though the Lyapunov exponents (and their differences)
are very small, eq. \eqref{eq:SALI-exp} still holds, as in the
strongly
chaotic domains of phase space \cite{kn:5}. %

If we slightly increase the energy, however, to $E=2.05$, SALI
displays an intermediate and rather intricate behavior, following a
stair--like decay to zero, shown by the middle curve in Fig.
\ref{fig:3}(c). Here, the exponential law \eqref{eq:SALI-exp} does
not explain the SALI decay to zero. It appears that the motion lies
within a weakly chaotic domain and {}``sticks'' temporarily to
islands of regular motion, executing a multi--stage diffusion
process \cite{kn:7}. Finally, by increasing the energy to $E=2.283$,
we observe that SALI decays to zero exponentially fast, following
closely the theoretical prediction
$e^{-(\lambda_{1}-\lambda_{2})t}$, with $\lambda_{1}\approx0.00166$
and $\lambda_{2}\approx0.0007$. These results imply that the
corresponding trajectory is fully chaotic and the system has
{}``melted'', passing to a regime with strongly mixing properties.

We now examine more closely this step--wise decay of the SALI, observed
in Fig. \ref{fig:3}(c), over a range of energies spanning the melting
transition. To this end, we start with our microplasma in the form
of a Wigner crystal, at $E_{0}$ and proceed to the onset of thermal
cloud formation, increasing the energy by steps of $\Delta E=0.05$
up to $E_{max}=6.2$. %
\begin{figure}[H]

\begin{centering}
\includegraphics[width=1\textwidth,height=0.5\textwidth]{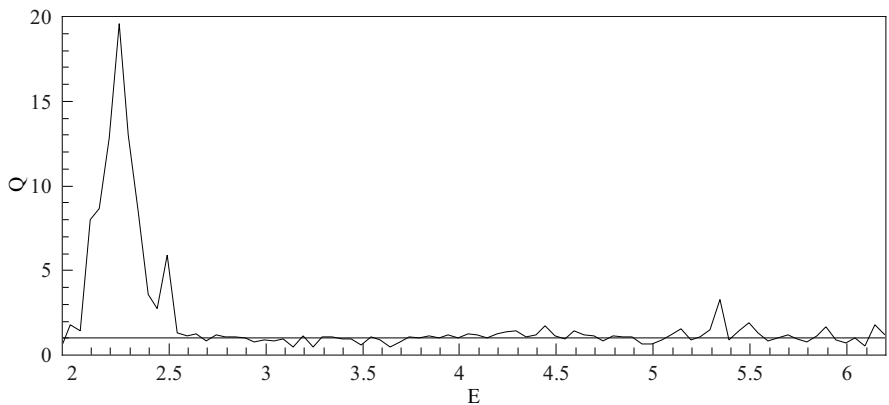}
\par\end{centering}

\caption{Plot of the quantity $Q=\frac{\Delta\lambda}{K}$, for $\gamma=0.07$ and $N=5$,  as a function of
the energy $E\in[E_{0},E_{max}]$ where $\Delta\lambda$ is the difference
of the two largest Lyapunov exponents and $K$ is the slope calculated
by linear regression performed on the $\log(\mbox{SALI}(t))$ vs.
$t$ curves. We also plot the line $Q=1$ to indicate the energy range
where $\Delta\lambda=\lambda_{1}-\lambda_{2}\approx K$ and eq. \eqref{eq:SALI-exp}
are satisfied. \label{fig:5} }

\end{figure}

One way to quantify the departure of the SALI from the exponential
decay law \eqref{eq:SALI-exp} (see Fig. \ref{fig:3}(c)) is to calculate
for each energy in the interval $[E_{0},E_{max}]$ the quantity \begin{equation}
Q=\frac{\Delta\lambda}{K}\label{Q-ratio}\end{equation}
 defined as the ratio of the difference between the two largest Lyapunov
exponents $\Delta\lambda=\lambda_{1}-\lambda_{2}$ and $K$, the
linear regression estimate of this difference, obtained from
semi--logarithmic plots of SALI, like those depicted in Fig.
\ref{fig:3}(b) and (c). The result is shown in Fig. \ref{fig:5}.
As expected, for the strongly chaotic cases where
$K\approx\Delta\lambda$, we obtain values of $Q$ near 1 with
clearly negative $\log$(SALI) slopes.

The interesting part of Fig. \ref{fig:5}, however, is observed in
the energy interval $2.0<E<2.5$ where $K$ becomes one order of
magnitude smaller than $\Delta\lambda$. It is in this regime that
we observe the ``melting transition'' for our system. The
divergence of SALI from its theoretical estimate of eq.
\eqref{eq:SALI-exp} signifies the presence of other collective
modes of motion. This motion is organized around islands of
stability, suggesting a certain type of ``stickiness'' around
different tori at successive time intervals. During these time
intervals SALI decays in a step--like manner by different power
laws.

In fact, we can provide more evidence to support the above
interpretation of the dynamics in the {}``melting phase'', using a
recent generalization of the SALI, called the \textit{Generalized
Alignment Indices} (GALI) \cite{kn:3}. These indices represent the
volumes of a parallelepiped, called GALI$_{k}$, formed by $k\geq2$
unit deviation vectors, which (a) for ordered motion, oscillate
about positive constants as long as $k\leq d$ ($d$ being the
dimension of the torus) and decay following power laws for $k>d$,
while (b) for chaotic orbits, they all vanish exponentially with
exponents that become more negative as $k$ increases and depend on
more Lyapunov exponents \cite{kn:3,kn:7}. In practise, since the
computation of the GALI requires the calculation of a fairly big
number of large determinants at every time step, we have adopted
an alternative way of evaluating it which is significantly faster
in CPU time, the so called \emph{Linear Dependence Index} (LDI)
\cite{kn:10}.

\begin{figure}[H]

\begin{centering}
\includegraphics[width=0.5\textwidth,height=0.5\textwidth]{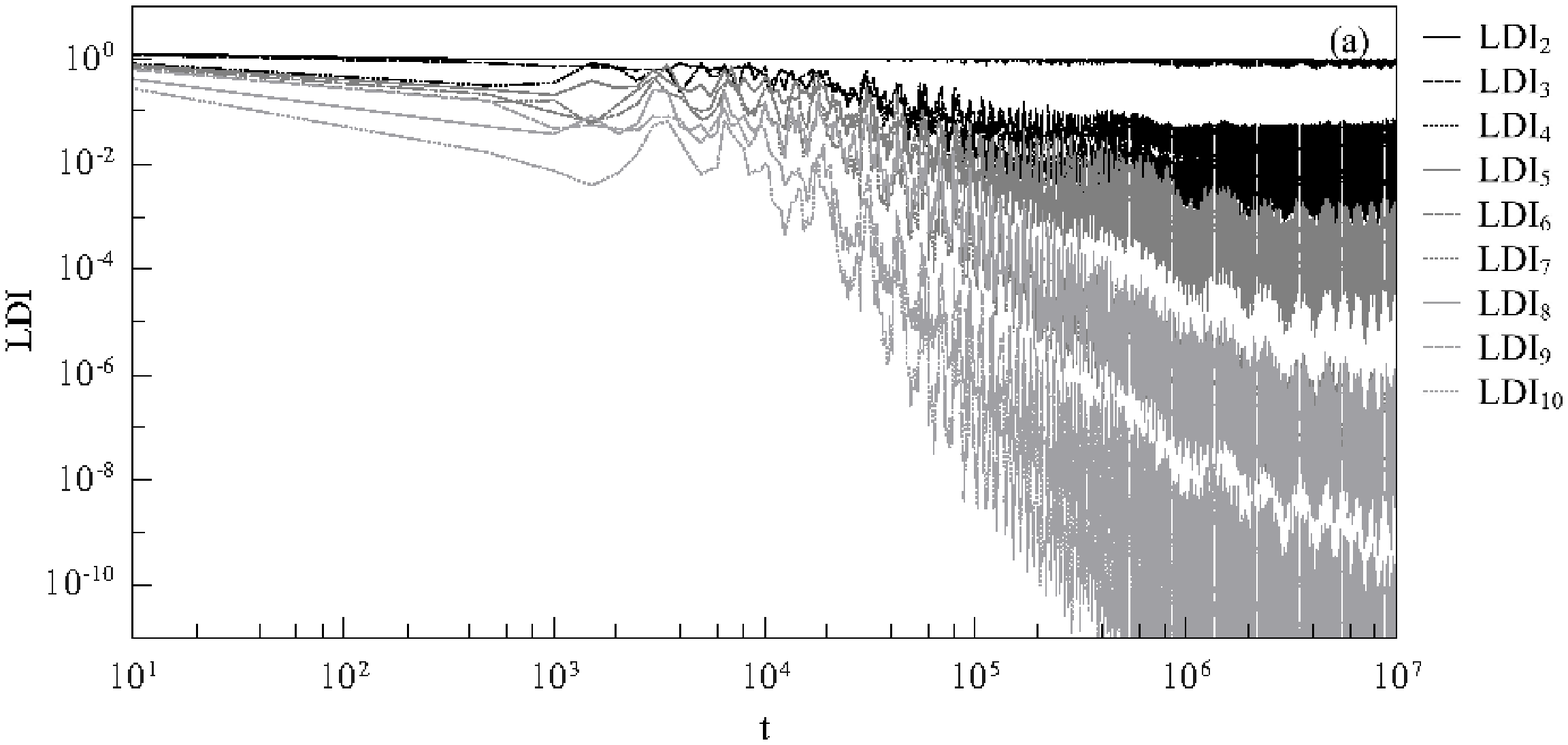}
\includegraphics[width=0.5\textwidth,height=0.5\textwidth]{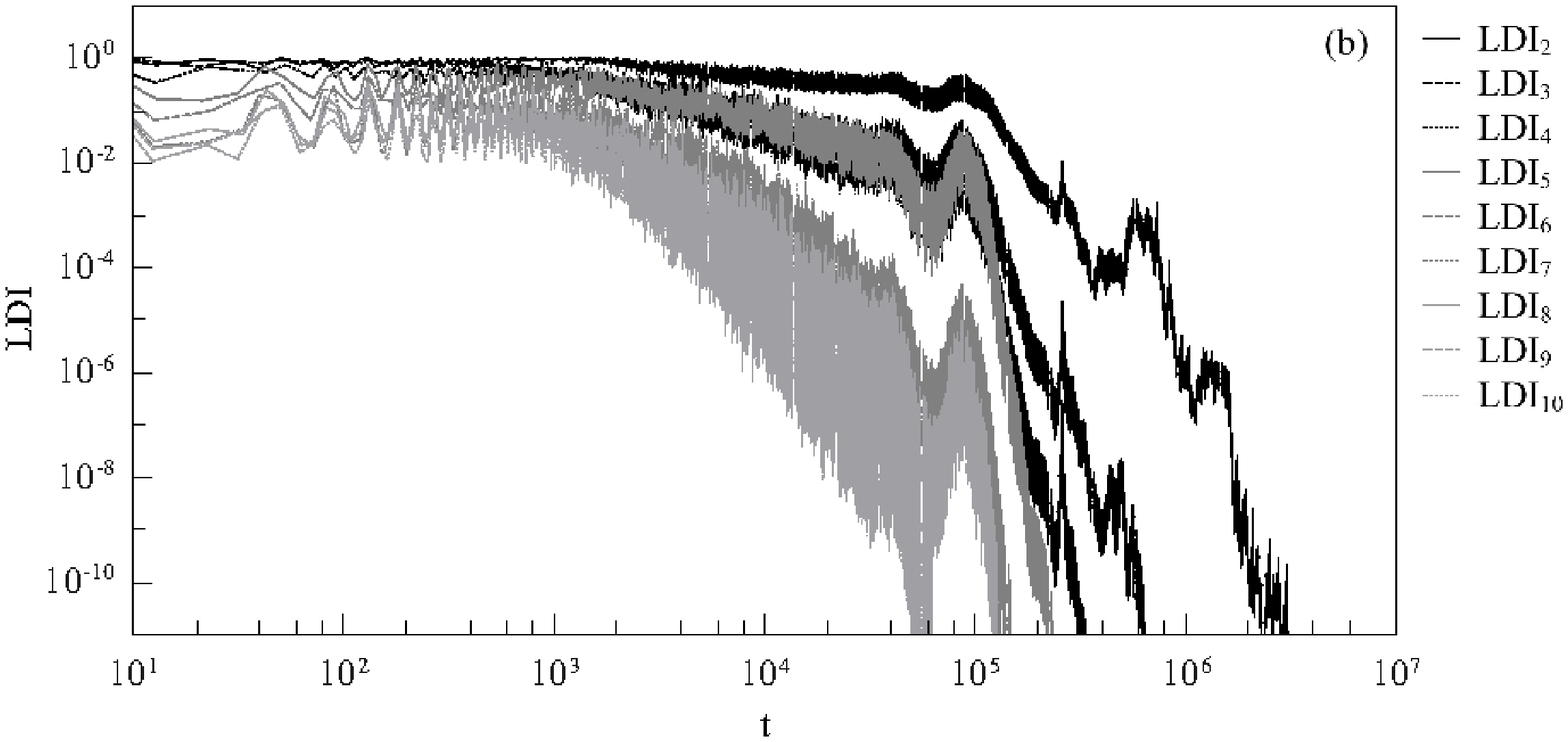}\includegraphics[width=0.5\textwidth,height=0.5\textwidth]{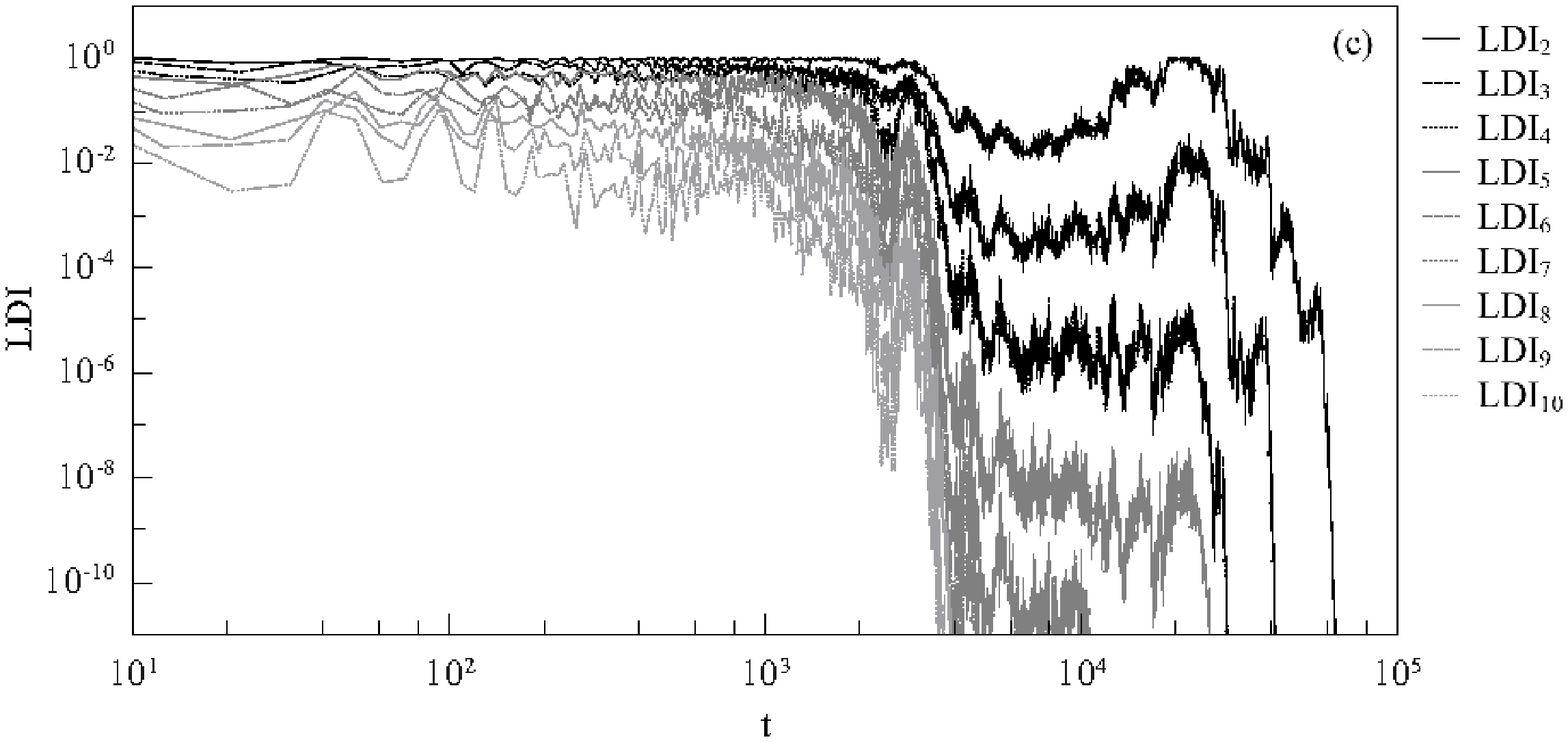}
\par\end{centering}

\caption{Plot of the LDI$_{k}$ (=GALI$_{k}$), $\; k=2,\ldots,10$ as a function
of time $t$ for the microplasma system \eqref{mic_plas_Ham} of $N=5$
ions in a prolate trap ($\gamma=0.07$) for 3 typical energies. (a)
LDI$_{2}$--LDI$_{10}$ for the energy $E=1.9$. (b) Same as in panel
(a) for the slightly bigger energy $E=2$. (c) Same as in panel (a)
for the bigger energy $E=2.283$. Note that all axes are logarithmic.\label{fig:6}}

\end{figure}

Thus, in Fig. \ref{fig:6}, we use this approach to compute the
GALI$_{k}$, $k=2,\ldots,10$ and find that they demonstrate the
existence of an (at least 6--dimensional) torus for $E=1.9$
\cite{kn:7}, while they show a step--wise decrease for $E=2$,
which continues to hold up to $E\lesssim2.283$, as predicted by
the SALI ($=\mbox{GALI}_{2}$) calculations (compare Fig.
\ref{fig:3}(a), (c) with Fig. \ref{fig:6}(a) and (b)). Of course,
as the energy increases, the (weakly) chaotic nature of the orbits
becomes evident at earlier times (e.g. for $E=2$ we have
$t\approx3\times10^{6}$ and for $E=2.283$ we get
$t\approx6\times10^{4}$ for the LDI (or GALI) to become
$\simeq10^{-11}$) as we see in Fig. \ref{fig:6}(b)
and (c). %

Thus, the above results suggest that the behavior of the SALI during
dynamical regime changes is especially useful, as it alerts us to
examine quantities like $Q$ (see eq. \eqref{Q-ratio}), related to
variations in the differences of Lyapunov exponents. Such quantities
indeed furnish important information, not readily available by other
more standard microscopic chaotic indicators or by related macroscopic
entropic quantities.

\section{Conclusions}

\label{concl} In this paper, we have reported results
demonstrating the occurrence of dynamical regime changes in a
Hamiltonian system describing a microplasma confined in a prolate
quasi 1--dimensional configuration and characterized by
long--range Coulomb interactions. More specifically, in the lower
energy regime, we macroscopically detected the transition from
{}``crystalline--like'' to {}``liquid--like'' behavior, through
what we call the {}``melting phase''. As expected, well beyond
this phase, the microplasma exhibits strong chaotic behavior that
may be described by the macroscopic variables of statistical
mechanics. The question, therefore, is how can one determine the
dynamical nature and identify the energy range of this {}``melting
transition'' from ordered to strongly chaotic behavior.

To this end, we first showed that the {}``melting'' of our quasi
1--dimensional crystal (composed of few ions confined in a prolate
trap) is not associated with a sharp increase of the temperature
at some critical energy, as might have been expected. Furthermore,
the positive Lyapunov exponents (and the Kolmogorov--Sinai entropy
expressed by their sum) attain their highest values at energies
much higher than the regime where the microplasma {}``melts''.
Thus, it appears that there is no clear {}``macroscopic''
methodology for identifying and studying this {}``melting''
process in detail.

Next, we argued that even though information about dynamical
regime changes must be contained in the Lyapunov exponents,
Lyapunov spectra by themselves still fail to provide, a more
precise picture of this change. Indeed, while the ratios of the
larger of them show strong fluctuations at low energies, these
require further analysis before they can reveal, with any
precision, the energies over which the {}``melting transition''
occurs.

For these reasons, we found it useful to employ a more accurate
tool, called the SALI method, to study in more detail the local
microscopic dynamics of the microplasma system. In this way, we
discovered an energy range of \textit{weak chaos}, where the
positive Lyapunov exponents are very small and SALI exhibits a
stair--like decay to zero with varying decay rates. As
$\mathrm{SALI}(t)\propto e^{-(\lambda_{1}-\lambda_{2})t}$ in
chaotic domains, this inspired us to look more closely at the
statistical fluctuations of the differences of the two largest
Lyapunov exponents $\Delta\lambda=\lambda_{1}-\lambda_{2}$ in that
regime. We, thus, observed that these differences exhibit their
largest fluctuations over a definite energy interval, where
``melting'' occurs in a way reminiscent of {}``sticky'' orbits,
executing a multi--stage diffusion process near the boundaries of
resonance islands in the phase space of Hamiltonian systems. The
above results were also supported by the use of an extended set of
indices which generalize SALI to the case of more than 2 deviation
vectors.

Finally, we remark that it is also the rapid convergence of these
indices, which turns them into efficient diagnostic tools that may
be used to replace demanding molecular simulations in identifying
weakly chaotic regimes in multi--particle systems. As we intend to
show in a future publication, these indices can provide a viable alternative
to the notoriously time consuming simulations required to study the
presence of weak chaos and slow diffusive effects in the dynamics
of metastable states.

\section{Acknowledgements}

The authors would like to thank professors G. Nicolis, P. Gaspard,
G--L. Forti and Ch. Skokos as well as Dr. P. de Buyl for fruitful
discussions, encouragement and support during the preparation of the
paper. We would also like to thank the referees for their constructive remarks and suggestions. 
The work of Ch. A. was financially supported by the PAI 2007--2011
``NOSY--Nonlinear systems, stochastic processes and statistical mechanics'' (FD9024CU1341) contract and the work 
of V. B. by the Prodex program of the European Space Agency under contract No. ESA AO--2004--070 (C90241). 
The computer programs of this work were ran in the high--performance multiprocessor
systems {}``PYTHIA'' of ULB and {}``TURING'' of UoP.

\bibliographystyle{plain} 

\end{document}